\documentstyle[preprint,aps,floats,epsfig]{revtex}

\newcommand{\be}{\begin{equation}}
\newcommand{\ee}{\end{equation}}
\newcommand{\bea}{\begin{eqnarray}}
\newcommand{\eea}{\end{eqnarray}}
\newcommand{\bml}{\begin{mathletters}}
\newcommand{\eml}{\end{mathletters}}
\newcommand{\lsim}{\mathrel{\raisebox{-.6ex}{$\stackrel{\textstyle<}{\sim}$}}}

\begin{document}
\preprint{hep-th/0012221}
\draft
\tighten

\title{  Living Inside the Horizon of the D3-Brane }
\author{ Won Suk Bae\footnote{E-mail: \texttt{ hypersym$@$phya.snu.ac.kr}},
         Y.M. Cho\footnote{E-mail: \texttt{ ymcho$@$yongmin.snu.ac.kr}} and 
         Sei-Hoon Moon\footnote{E-mail: \texttt{ jeollo$@$phya.snu.ac.kr}}}

\address{School of Physics, Seoul National University, Seoul 151-742, Korea }
\date{\today}
\setlength{\footnotesep}{0.5\footnotesep}
\maketitle

\begin{abstract}
We consider a brane world residing in the interior region inside the horizon 
of the D3-brane. The horizon size can be interpreted as the compactification 
size. The macroscopically large size of extra dimensions then can be derived 
from the underlying string theory that has only one physical scale, {\it i.e.,}
the string scale. Then, the hierarchy between the string scale and the Planck
scale is provided by Ramon-Ramon charge of the D3-brane. This picture also
offers a new perspective on various issues associated with the brane world 
scenarios including the cosmological constant.
\end{abstract}


\newpage

\section{Introduction}

String theory spacetimes with conserved quantum numbers are commonly black 
$p$-branes \cite{HS}. In particular, the extremal $p$-brane with R-R charge 
can be described as a D$p$-brane \cite{Pol}. The discovery of D-brane 
drastically changed the conventional viewpoint that the standard model fields 
and the graviton should propagate in the same spacetime. The standard model 
fields can be naturally confined on D-brane, while the graviton can freely 
propagate through the full spacetime. Recently, this concept has been 
extensively introduced in approaches to resolving the hierarchy problem 
and getting a small cosmological constant without relying on supersymmetry
in the framework of brane world scenarios. 

In Refs.\cite{ADD,AADD,STR1,STR2}, it was shown that the string scale 
$l_s^{-1}$ could be reduced from the four-dimensional Planck scale, 
$M_{pl}\sim 10^{19}$GeV, to the electroweak scale, $m_{EW}\sim 10^3$GeV, by 
introducing large extra dimensions. However, this scenario has the problem of 
stabilization of the large extra dimensions or, equivalently, dynamical 
determination of the large size of extra dimensions in a theory whose 
fundamental length is the inverse weak scale. 

Subsequently, Randall and Sundrum (RS) proposed a scheme in which a 
five-dimensional spacetime contains strongly gravitating 3-branes which 
produce a warped or nonfactorizable geometry, {\it e.g.} a slice of $AdS_5$. 
In their first model (RSI) with two 3-branes \cite{RS1}, solving the hierarchy 
problem does not require large extra dimensions. In the second model (RSII) 
\cite{RS2} which has only one 3-brane of positive tension, RS also showed 
that, even for an infinite extra dimension, the effective gravity theory on 
the brane can be still four-dimensional and not five-dimensional up to leading 
behavior, provided that the embedding spacetime has a very particular 
curvature. 

Many attempts to resolve the cosmological constant problem have been made 
within the brane world scenarios \cite{RuSh,ADKS,KSS,CEGH,FLLN,Luty,CC,Al}: 
If our four-dimensional world 
is embedded in a higher-dimensional spacetime, the changes of the vacuum 
energy of the brane, the brane tension, may affect only the curvature in the 
extra dimensions, keeping the Poincar\'{e} invariance of the four-dimensional 
worldvolume.     

Various attempts have been made to realize the Randall-Sundrum scenario within 
5-dimensional supergravity theories \cite{CLP,CLP1,BDPS,DLS,KL} and a 
compactified string or M-theory \cite{BER,VV,CPV}. In particular, the authors 
of Ref.\cite{CLP} investigated singular, supersymmetric domain-wall solutions 
supported by the massive breathing mode scalars of reductions in M-theory or 
string theory. The spacetime on one side of such a wall asymptotes to the 
Cauchy horizon of the anti-de Sitter (AdS) spacetime, while there is a naked 
singularity on the other side. The higher dimensional interpretation of these 
domain wall solutions is given as the interior region between the singularity 
and the horizon of the higher-dimensional non-dilatonic $p$-branes 
\cite{CLP,BDPS}. This implies that the interior region could provide a setup 
for brane world scenarios, even though it is normally excluded from the maximal 
analytic extension.

On the other hand, the authors of Ref.\cite{KMR} considered an extremal black 
hole-like global defect which was called as `black global $p$-brane', and 
showed that the interior region inside the horizon of the black brane possesses
all of features needed for a realistic brane world scenario. In the picture of 
Ref.\cite{KMR}, the size of the horizon can be interpreted as the 
compactification size, even though the interior region inside the horizon 
infinitely extends.
And the large mass hierarchy is generated from topological charge 
of the brane, from which the horizon size is determined.
In this picture, the change of the brane tension can forcefully be diluted 
via the Hawking radiation, because the change of the brane tension convert 
the bulk geometry of the extremal black brane into that of a non-extremal one, 
which evolves back to the extremal one through the Hawking radiation process.  

In this paper, we consider the D3-brane living in the ten-dimensional 
spacetime and focus on the interior region inside the horizon of the D3-brane.
The interior region has interpolation between the naked singularity and the 
near horizon region which is the infinitely long AdS throat ($AdS_5\times S^5$).
If the singularity is smoothed out by stringy effects, the physics will be
close to that of the global black brane of Ref.\cite{KMR}. The interior region 
of D3-brane then can be interpreted as four-dimensional spacetime producted by 
six extra dimensions compactified to the size of the horizon, in that the 
four-dimensional Planck scale $M_{pl}$ is determined by the 
ten-dimensional Planck scale $M_*\sim g_s^{-1/4}l_s^{-1}$ and the horizon size 
$r_H$ via the relation $M_{pl}^2\sim M_*^8r_H^6$, and the gravity on a 3-brane
residing in the interior region behaves as expected in a world with six extra 
dimensions compactified to size $r_H$. As the result of such interpretation,
the compactification scale can be dynamically determined from the underlying
string theory which has only one physical scale, {\it i.e.}, string scale 
$l_s^{-1}$. A large mass hierarchy is then simply translated into large
horizon size. Moreover, since the size of the horizon is determined by the 
Ramon-Ramon (R-R) charge of D3-brane, the large mass hierarchy then follows by 
taking large R-R charge and its stabilization is guaranteed by the  
conserved nature of the R-R charge. The observed extremely small cosmological 
constant could be explained provided that our world brane is embedded in the
interior region of a very near-extremal D3-brane.    

\section{The spacetime of D3-brane}

In string theory, the D$p$-brane is a $(p+1)$-dimensional hyperplane in 
spacetime where open strings can end \cite{Pol}. By the worldsheet duality, 
this means that the D-brane is also a source of closed strings. If there are 
$N$ overlapping branes, open strings with endpoints on different branes 
correspond to $U(N)$ gauge fields localized in the world-volume of D$p$-brane. 
On the other hand, gravity corresponds to closed strings and these can move on 
the whole 10-dimensional spacetime. 

On the other hand, it is believed that the D$p$-brane and the extremal black
$p$-brane in supergravity are two different descriptions of the same object.
For D$p$-branes with $p\neq3$, the horizon and the singularity coincide, so
there is null singularity. Moreover, the dilaton either diverges or vanishes 
at the horizon. 
While, for the D3-brane, the dilaton is constant and the horizon is regular. 
Together with these properties, since the D3-brane has a $(3+1)$-dimensional 
world volume, it may provide an interesting and concrete model for attempts
to describe our universe as a 3-brane embedded in a higher-dimensional 
spacetime. 

The supergravity solution of D3-brane with $N$ units of R-R charge is given by 
\begin{equation}\label{ebs}
ds^2=\sqrt{\sigma\Delta(r)}~\eta_{\mu\nu}dx^\mu dx^\nu
     +\Delta(r)^{-2}~dr^2 + r^2d\Omega_5^2,
\end{equation}
in terms of an appropriately defined Schwarzschild-type coordinate $r$, where 
\begin{equation}
\Delta(r)\equiv1-\left(\frac{r_H}{r}\right)^4
~~~{\rm with}~~r_H^4\equiv 4\pi l_s^4g_sN,
\end{equation}
where $l_s$ is the string length scale and $g_s$ is the string coupling. 
Here $\sigma=-1$ for the interior solution ($r<r_H$) and $\sigma=1$ 
for the exterior solution ($r>r_H$). The spacetime has a horizon at $r=r_H$ 
and this horizon is regular. And the singularity at $r=0$ is covered with the 
horizon. As discussed in \cite{GHT}, it is known that the metric (\ref{ebs}) 
can be analytically extended through the Cauchy horizon at $r=r_H$ and out 
into another region.
This can be seen by introducing a new radial coordinate $\omega$ by the relation
\begin{equation}
\omega=[\sigma\Delta(r)]^{1/4}~~\Leftrightarrow~~
r=r_H\left(1-\sigma\omega^{4}\right)^{-1/4}. 
\end{equation}
Since $r$ is an analytic function of $\omega$ on the horizon at $\omega=0$, the 
metric can be extended to negative values of $\omega$. 
The extension is isometric to the original regions because the metric is 
invariant under $\omega\rightarrow -\omega$. For the exterior solution, 
repeating the continuation, one obtains a completely non-singular maximal 
analytic extension of the metric (\ref{ebs}). The interior region is excluded 
from this extension. This is the case normally treated in the literatures. 
However, we have no reason to exclude the interior region from discussions 
given our ignorance concerning the singularity. In fact, there must be a 
R-R charged $\delta$-function source at $r=0$ because the D3-brane is 
coupled to the self-dual 5-form field strength and carry both \lq\lq electric" 
and \lq\lq magnetic" charge. The interior region then can be also analytically 
extended through the horizon to another region isometric to the original 
interior region excluding the exterior region. 
 
As well known, the near horizon geometry is the infinitely long AdS throat. 
To see this, it will be useful to introduce a new radial coordinate $\rho$ by
\begin{eqnarray}\label{rho1}
\rho&=&\int^r\left[\sigma\Delta(r')\right]^{-5/4}~dr' ,
\end{eqnarray}
where $\rho$ runs from $0$ to $\infty$ for the interior solution and from 
$-\infty$ to $\infty$ for the exterior solution. The horizon is now located at 
$\rho=\infty$ for the interior solution and at $\rho=-\infty$ for the exterior
solution. With this new radial coordinate, the metric is then written as
\begin{equation}
ds^2=\sqrt{\sigma\Delta(r(\rho))}~\left(\eta_{\mu\nu}dx^\mu dx^\nu
      +d\rho^2\right)+r(\rho)^2 d\Omega_5^2 ,
\end{equation}
where the original radial coordinate $r$ is given by a function of $\rho$ and 
its asymptotic behaviors can easily be found. In the near horizon region, we 
can approximate the metric as 
\begin{equation}
ds^2\approx\frac{r_H^2}{\rho^2}\left(\eta_{\mu\nu}dx^\mu dx^\nu
        +d\rho^2\right)+r_H^2 d\Omega_5^2 ,
\end{equation}
which is the geometry of $AdS_5\times S^5$ with curvature scales $r_H^{-1}$.
This solution is an example of warped metric due to the backreaction of the
brane. 

So far the black 3-brane have been treated by using the classical supergravity.
This description is appropriate when the curvature of the 3-brane geometry
is small compared to the string scale, so that the stringy corrections
are negligible. For the exterior region, since the curvature is bounded
by $r_H^{-1}$, this requires $r_H\gg l_s$. To suppress string loop corrections,
the effective string coupling $g_s$ also needs to be kept small. 
Since the dilaton is constant, we can make it small everywhere in the D3-brane 
geometry by setting $g_s<1$, namely $l_P<l_s$. 
That is, the supergravity approximation is valid everywhere when 
$l_p<l_s\ll r_H$ for the exterior solution. Since $r_H$ is related to the R-R 
charge $N$ as $r_H^{4}=4\pi l_s^4 g_sN$, this can also be expressed as 
$1\ll g_sN<N$. On the other hand, since in the interior region the curvature 
is not bounded at the origin, the supergravity description will be valid only 
in a limited region. Because the curvature diverges as $\sim r_H^4/r^5$ near 
the singularity, the interior solution is valid within the region satisfying 
the condition $r\gg (4\pi g_sN)^{1/5}l_s$ and the stringy description might be 
needed in a region where $r\lsim (4\pi g_sN)^{1/5}l_s$. In terms of the radial 
coordinate $\rho$, this corresponds to the region where 
$\rho\lsim \left(4\pi g_sN\right)^{-1/20}l_s\lsim l_s$.
This implies that the singularity could be an artifact of the 
long-range supergravity approximation. The singularity is expected to be 
cured by stringy effects. 
Essentially the source for the R-R field strength is sitting
at the singularity, {\it i.e.,} the field energy density diverges as 
$\sim 1/r^{10}$. Since only the known source for the R-R 5-form field 
strength is D3-brane, we are naturally expected to see a stack of D3-branes 
when we probe the singularity with energy over the string scale.  

\section{Metric fluctuations and the singularity}

Either the interior or the exterior region by itself does not seem to provide
a suitable framework for a Randall-Sundrum configuration, because in one 
direction one may reach an AdS horizon, in the other direction one will either
run into a singularity or on out into an unbounded flat space. 
However, as discussed in \cite{CLP} in the context of the five-dimensional 
domain wall solution to the dimensionally reduced type IIB theory on $S^5$, 
such singular region by itself can provide a suitable framework for a 
Randall-Sundrum configuration, provided that the singularity is regularized.   
  
The existence of the massless graviton state is evident because the theory 
always allows solutions with a general Ricci-flat metric $\bar{g}_{\mu\nu}(x)$, 
which satisfies the four-dimensional vacuum Einstein equation 
$\bar{R}_{\mu\nu}(\bar{g})=0$, instead of $\eta_{\mu\nu}$ 
in the metric (\ref{ebs}), and the massless four-dimensional graviton is simply
the usual gravitational wave solution of linearized four-dimensional vacuum 
Einstein equation. The Planck scale is finite in spite of the 
existence of the singularity in the interior region. One can easily see this
examining the four-dimensional effective action. The four-dimensional 
Planck scale is given in terms of the 10-dimensional Planck scale by:
\begin{equation}\label{psl}
M_{pl}^2=M_*^8\int d\Omega_5\int_0^{r_H}dr~r^5[-\Delta(r)]^{-1/2}
        = M_*^8\cdot\frac{\pi^3}{3} r_H^6,
\end{equation} 
where $M_*$ $(\equiv (8\pi^6g_s^2l_s^8)^{-1/8})$ is the 10-dimensional Planck 
scale. $(\pi^3/3)r_H^6$ is just $5/8$ times the volume of 6-dimensional sphere,
so the radial direction has the effective size of $(5/8)r_H$. 

Eq.(\ref{psl}) says that the four-dimensional Planck scale is determined by 
the ten-dimensional Planck scale and the horizon size via the familiar
relation $M_{pl}^2\sim M_*^8 r_H^6$ as in the usual Kaluza-Klein theories. 
This implies that the horizon size $r_H$ could be interpreted as the effective 
size of 6 compact extra dimensions, even though the interior region infinitely 
extends\footnote{The six extra dimensions are a direct product of 
five-dimensional sphere $S^5$ with radius $r_H$ and an infinitely extended 
radial direction. Despite being noncompact, due to the warped spacetime 
geometry, the radial direction yields a finite effective size $\sim r_H$.}. 
This interpretation could be cleared up through a complete 
analysis of the effective 4D gravity.

However, we should be careful in discussion of the graviton state because we 
do not have a complete understanding of the singularity. For a metric of the 
form of Eq.(\ref{ebs}), the gravitational fluctuations are given by replacing 
$\eta_{\mu\nu}$ with $\eta_{\mu\nu}+h_{\mu\nu}(x,y)$ where
$h_{\mu\nu}$ is small compared with $\eta_{\mu\nu}$. 
With the gauge condition $\partial^\mu h_{\mu\nu}=0$, the linear 
fluctuations satisfy the covariant wave equation\cite{CEHS}:
\begin{equation}\label{we}
\frac{1}{\sqrt{g}}\partial_M\left(\sqrt{g}~g^{MN}\partial_Nh_{\mu\nu}\right)=0,
\end{equation}  
where $g_{MN}$ is the ten-dimensional background metric. 
Using the coordinates of the metric (\ref{ebs}) and expanding $h_{\mu\nu}
(x,r,\Omega)=\epsilon_{\mu\nu}e^{ip\cdot x}R_{\ell}(r)Y_{\ell}(\Omega)$, 
where $\Omega$ denotes angular variables, we have an equation for the radial
wavefunction $R_{m\ell}(r)$: 
\begin{equation}\label{le}
-\frac{1}{r^5}\frac{d}{dr}\left(\Delta(r)^2~r^5\frac{d}{dr}\right)R_{m\ell}(r)
+\frac{\ell(\ell+4)}{r^2}R_{m\ell}(r)=m^2R_{m\ell}(r),
\end{equation}
where $\epsilon_{\mu\nu}$ is a constant polarization tensor and $m$ 
$(=\sqrt{-p\cdot p})$ is the mass of the KK modes. The second term is the 
repulsive centrifugal potential which comes from the contribution of angular 
momentum. As easily expected, this equation is singular at $r=0$.
Since this equation has the asymptotic form near the singularity:
\begin{equation}
-\frac{r_H^8}{r^8}\frac{d^2R_{m\ell}}{dr^2}+3\frac{r_H^8}{r^9}
\frac{dR_{m\ell}}{dr}+\ell(\ell+4)\frac{R_{m\ell}}{r^2}=m^2R_{m\ell},
\end{equation} 
the equation (\ref{le}) requires regular solutions to behave near the 
singularity as follows
\begin{eqnarray}\label{bc}
&& R_{m0}(r)\rightarrow a_0\left(1-\frac{m^2}{60r_H^8}r^{10}\right)
             +\cdot\cdot\cdot,~~~~{\rm for}~\ell=0, \\  
&& R_{m\ell}(r)\rightarrow r^n~~{\rm with}~~n>10,
             ~~~~~~~~~~~~~{\rm for}~\ell\neq0, 
\end{eqnarray}
where $a_0$ is a constant.
In the presence of the singularity, a possible boundary condition is simply 
to require that the fields should not feel the singularity at all: 
$R_{m\ell}(0)=0$ disallowing solutions with $R_{m\ell}(0)\neq 0$ at $r=0$ 
as in \cite{CLP}. 
Then the spectrum is bounded from below and 
continuous with only positive energies occurring\cite{DFGK,CEHS}. 
And the wavefunctions with $m\neq0$ will be suppressed in most of the interior 
region, as will be cleared up later. 
For the massless graviton ($i.e.,~m=\ell=0$), Eq.(\ref{le}) admits the constant 
solution $R_{00}(r)=$ constant. However, this massless graviton state should 
be excluded, since it does not vanish at $r=0$. Hence, the inverse square law 
for gravity will not be recovered at long distance on a brane residing in the 
interior region. 

Another possible boundary condition is the unitary boundary condition which was 
discussed in \cite{CK,grem}.
The fact that the spacetime is geodesically incomplete would not matter provided
that no conserved quantities are allowed to leak out through the boundary.
Even though the singularity of the black 3-brane at $r=0$ is point-like and 
stronger than those considered in \cite{CK,grem}, it is easy to see, following 
the same procedure as in Refs.\cite{CK,grem}, that such 
boundary condition allows the form of solution (\ref{bc}), that is, the flux 
through the singularity for the solutions having behaviors of (\ref{bc}) 
vanishes. Further, the constant solution for the massless graviton is not 
excluded by such boundary condition. 
Hence, the usual four-dimensional gravity will be reproduced at long distance, 
even though the spacetime has a naked singularity.  

However, since we do not have a good understanding of the singularity,
we will not be able to make any rigorous claim whether such boundary 
conditions are what we want. Thus, we will simply assume that the singularity 
is smoothed out by the true short-distance theory of gravity, namely, string 
theory. The singular source of the R-R field then will be look like a smooth 
3-brane.  Since these effects will only modify the metric close to the 
singularity, they only slightly change the shape of the wavefunctions.
The massless graviton then is always allowed and the correct four-dimensional 
gravity is reproduced at long distance on a the 3-brane. 

\section{Gravity on the brane and mass hierarchy}

It will be useful to work using the radial coordinate $\rho$ of Eq.(\ref{rho1})
to see the effective gravity on a 3-brane residing in the interior region. 
The linearized equation (\ref{le}) then can be written into the form of an 
analog non-relativistic Scar\"{o}dinger equation by making a change of variable
\begin{eqnarray}
\chi_{m\ell}(\rho)=K(\rho)~R_{m\ell}(r(\rho)),
\end{eqnarray}
where
\begin{equation}\label{kro}
K(\rho)\equiv \left(\frac{r(\rho)}{r_H}\right)^{5/2}
              \left[-\Delta(r(\rho)) \right]^{3/8}
  \simeq\left\{\begin{array}{ll}
                  \left(\displaystyle{\frac{6\rho}{r_H}}\right)^{1/6} 
                    &~~~{\rm for}~~\rho\ll r_H \\
                  \left(\displaystyle{\frac{\rho}{r_H}}\right)^{-3/2}
                    &~~~{\rm for}~~\rho\gg r_H
               \end{array}\right. .
\end{equation}
Then the resulting analog Schr\"{o}dinger equation is given by
\begin{equation}\label{sch}
\left[-\frac{\partial^2}{\partial\rho^2}+V_{sch}(\rho)+V_{cfl}(\rho)\right]
\chi_{m\ell}(\rho) =m^2~\chi_{m\ell}(\rho),
\end{equation}
where $V_{sch}(\rho)$ is the analog non-relativistic quantum mechanical 
potential:
\begin{eqnarray}\label{vsch}
V_{sch}(\rho)\equiv \frac{K''(\rho)}{K(\rho)}
             &=&\frac{5}{4r_H^2}\frac{3y(\rho)^8+4y(\rho)^4-4}{y(\rho)^{10}}
                \sqrt{\frac{1}{y(\rho)^4}-1} \\
             &\simeq& 
             \left\{\begin{array}{ll}
                  -\displaystyle{\frac{5}{36}\frac{1}{\rho^2} }
                    &~~~{\rm for}~~\rho\ll r_H \\
                  \displaystyle{\frac{15}{4}\frac{1}{\rho^2} }
                    &~~~{\rm for}~~\rho\gg r_H
               \end{array}\right.  , 
\end{eqnarray}
where $y(\rho)\equiv r(\rho)/r_H$, 
and $V_{cfl}$ is the repulsive centrifugal potential:
\begin{eqnarray}\label{vcfl}
V_{cfl}(\rho)\equiv\frac{\ell(\ell+4)}{r_H^2}\frac{[y(\rho)^{-4}-1]^{1/2}}
             {y(\rho)^2}\simeq\left\{\begin{array}{ll}
 \displaystyle{\frac{\ell(\ell+4)}{r_H^2}\left(\frac{r_H}{6\rho}\right)^{2/3}} 
                    &~~~{\rm for}~~\rho\ll r_H \\
              \displaystyle{ \frac{\ell(\ell+4)}{\rho^2} }
                    &~~~{\rm for}~~\rho\gg r_H
               \end{array}\right.     .
\end{eqnarray}
\begin{figure}[htbp]
\centerline{\epsfig{file=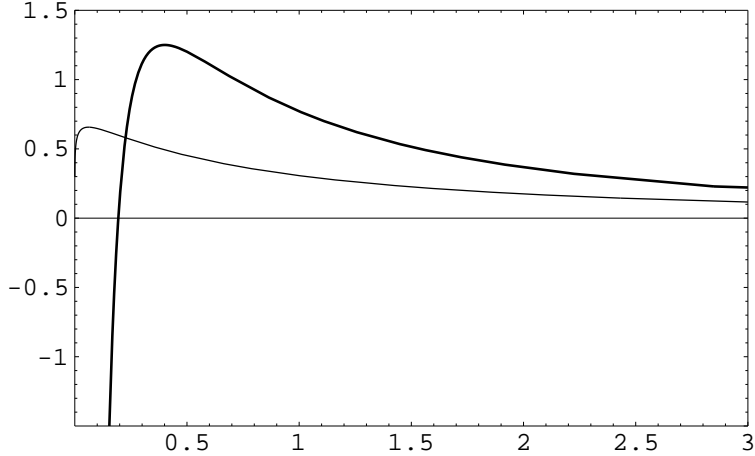,width=10cm}}
\caption{
Plots of the Schr\"{o}dinger potential $r_H^2\cdot V_{sch}$ (thick solid line) 
and the massless graviton mode $K(\rho)$ (thin solid line) as functions of 
$\rho/r_H$.
}
\label{fig:schpo}
\end{figure}
All of the important physics follows from a qualitative analysis of these
potentials. The full form of the Schr\"{o}dinger potential $V_{sch}(\rho)$ is 
sketched in FIG. \ref{fig:schpo}. 
We see that the potential near the singularity approaches a negative infinity
as $\sim -\rho^{-2}$. However, the negative infinity of the potential should be 
cut off at distances $\rho\sim l_s$ after smoothing out the singularity.
Without a detailed understanding of the theory in such 
strongly gravitating region, we will not know the shape of the potential and 
so the shape of the wavefunctions near the origin. In \cite{CLP}, a model 
which modify the metric within the distance $\rho\sim l_s$ was presented to 
get an indication of the sort of modifications that can be expected once 
stringy corrections are taken into account. 
As we will see, however, such models are not needed in analyzing the 
wavefunctions in the interior region, because the amplitude of the wavefunction 
near the origin is insensitive to the detailed shape of the potential near the 
origin. 

A bump results from the AdS spacetime near the horizon. The 
potential has a maximum value $5/4r_H^2$ at $\rho\approx0.40r_H$ and is zero 
at $\rho\approx 0.19r_H$ showing that the central region of the potential is 
smaller than the AdS scale $r_H^{-1}$. The potential also approaches zero as 
$\sim \rho^{-2}$  near the horizon at $\rho=\infty$. For $s$-wave modes with 
$\ell=0$, the potential in the near horizon region is identical to that of the 
original Randall-Sundrum model. While wave modes with $\ell\neq0$ will see the
additional repulsive centrifugal potential and will be much more suppressed 
near the origin than $s$-wave modes. 

The zero mode solution is easily identified from Eq.(\ref{sch}) as 
$\chi_{00}(\rho)=a_{00}K(\rho)$, which corresponds to the constant zero-mode 
solution of Eq.(\ref{le}). Here, $a_{00}$ ($\equiv(\pi^3r_H^6/3)^{-1/2}$) is 
the normalization factor. The function $K(\rho)$ is plotted
in FIG.\ref{fig:schpo}. Clearly, the graviton is localized in the central 
region of the potential, even though $K(\rho)$ rapidly approaches zero as
$\sim(\rho/r_H)^{1/6}$ at the very near of the origin when the regularization 
is not imposed. This seems to show that the gravity is completely decoupled 
from the $\delta$-function like source at the singularity. The function $K(r)$, 
however, should be modified near the origin when the singularity is regularized.
Since $K(l_s)\sim (l_s/r_H)^{1/6}$ at $\rho\sim l_s$, the function $K(r)$
does not significantly change in the central region $l_s\lsim\rho\lsim0.4r_H$,
even for the large difference of order $l_s/r_H\sim 10^{-5}$.
Hence, the central region looks like a one-sided Randall-Sundrum domain wall 
embedded in $AdS_5$ upon dimensional reduction.  
And the continuum modes in the near horizon region are given by a linear 
combination of Bessel functions
\begin{equation}\label{bes}
\chi_{m\ell}(\rho)=\sqrt{\rho}\left[ a_{m\ell}Y_{2+\ell}(m\rho)+
                      b_{m\ell}J_{2+\ell}(m\rho)\right],
\end{equation}
where $a_{m\ell}$ and $b_{m\ell}$ are $m$-dependent coefficients to be 
determined by normalization and boundary conditions. 
The bump resulting from the AdS spacetime causes the continuum modes with 
masses $m\ll r_H^{-1}$ to be suppressed in the central region of the potential 
(close to the singularity). It is difficult to find quantitative 
behaviors of the continuum modes in the central region of the potential 
directly solving the Schr\"{o}dinger equation. 
However, when the central region of the potential is localized within the 
AdS curvature length scale, the behaviors of the continuum modes can be 
determined from the asymptotic solution (\ref{bes}) using some matching 
conditions, as discussed in \cite{CEHS}. 
Following the same procedure as 
in \cite{CEHS}, we obtain the value of the radial wavefunction in the central 
region up to the leading order in $m$,
\begin{equation}\label{chi0}
\chi_{m\ell}(\rho)\sim\left(\frac{m}{r_H^{-1}}\right)^{\ell+1/2}. 
\end{equation}
While the continuum modes with $m>r_H^{-1}$ will sail over the potential and
will be unsuppressed in the central region. 
In order to see the physics more explicitly, suppose a stack of D3-brane 
sitting at the origin which is strongly gravitating or a probe 3-brane residing 
at a point in the central region $\l_s\lsim\rho\lsim0.4r_H$ which is very 
light. The gravitational potential $U(|\vec{x}|)$ of a test particle with mass 
$m^*$ on the brane then is given by
\begin{eqnarray}\label{grv}
\frac{U(|\vec{x}|)}{m^*}
        \sim G_{4}\frac{1}{|\vec{x}|}+ \frac{1}{M_*^8}\sum_{\ell}
     \int_{m\neq0}\frac{dm}{r_H^{-1}}\left(\frac{m}{r_H^{-1}}\right)^\delta 
           |\chi_{m\ell}(\rho)|^2 \frac{e^{-m|\vec{x}|}}{|\vec{x}|},
\end{eqnarray} 
where $G_{4}$ ($\sim |\chi_{00}(\rho)|^2/M_*^8$) is the four-dimensional 
gravitational constant and $|\vec{x}|$ is the distance from the test particle 
on the brane. $\delta$ is a power of proper measure for sums over the 
continuum modes and has $\sim0$ for small $\ell$ and $\sim5$ for large $\ell$.
For large distances $|\vec{x}|\gg r_H$, the first term generated by the 
four-dimensional graviton bound state is dominant and, using Eq.(\ref{chi0}),
we get 
\begin{equation}
U(|\vec{x}|) \sim G_{4}\frac{m^*}{|\vec{x}|}\left[1+\frac12\left(\frac{r_H}
{|\vec{x}|}\right)^2+\sum_{\ell\neq0}\left(\frac{r_H}{|\vec{x}|}
\right)^{\delta+2+2\ell}\right].
\end{equation}
Since modes with $m\ll r_H^{-1}$ only contribute to the long distant gravity, 
modes with small $\ell$ are dominant and so $\delta\sim0$. The leading 
correction behave like $(r_H/|\vec{x}|)^2$, as in the original Randall-Sundrum 
model, rather than $(r_H/|\vec{x}|)^7$, which may usually be expected to be 
for Randall-Sundrum type models with $6$ extra dimensions.
On the other hand, for distances $|\vec{x}|\ll r_H$, the continuum modes with
$m\gg r_H^{-1}$ are dominant and has unsuppressed wave functions in the central 
region. Modes with large $\ell$ largely contribute to the short distance 
gravity and then $\delta\sim5$ in the integrand of Eq.(\ref{grv}). 
Therefore, for $|\vec{x}|\ll r_H$, we get the 10-dimensional gravitational 
potential
\begin{equation}
U(|\vec{x}|) \sim \frac{1}{M_*^8}\frac{m^*}{|\vec{x}|^7}.
\end{equation}  

Hence, the behavior of the gravity on the brane is exactly what we would 
expect by interpreting $r_H$ as a \lq\lq compactification radius", concretely
supporting the interpretation of the horizon size as the compactification 
scale. That is, the interior region of the D3-brane looks like the 
four-dimensional spacetime producted by six extra dimensions compactified to
the size of the horizon. 
Indeed, to an observer living outside the horizon, this interpretation 
seems correct in that the interior region occupies only a finite part of 
volume $\sim r_H^6$ of the transverse space. The apparent infinite extension
of the interior region is simply the result of the warping of the extra 
dimensions by the gravity produced by the D3-brane. 

The physics on a brane residing in the central region interpolates between the 
Randall-Sundrum one and the large extra dimensional scenario relying on the 
size $r_H$ of the horizon. In case that the string scale is of the order of 
the four-dimensional Planck scale, {\it i.e.,} $l_s^{-1}\sim M_{pl}$, the size 
of the horizon is to be of the order of the string scale. Thus, at low energy
below the Planck scale, the extra space is reduced effectively to a 
one-dimensional space. Consequently, the central region looks like a one-sided 
Randall-Sundrum brane embedded in $AdS_5$ bulk spacetime. 
Since the information on the compactified sphere $S^5$ is reflected only 
through the extremely suppressed contribution from modes with $\ell\neq0$ and 
moduli fields associated with the radius $r_H$ of $S^5$ have masses of the 
order of the string scale, the phenomenology on the brane world is 
imperceptibly different from that of the original Randall-Sundrum scenario. 
This is what is expected from the supergravity domain wall solution
of Ref.\cite{CLP} to the dimensionally reduced type IIB theory on $S^5$.
However, this limit does not seem to be reliable because the supergravity 
description of the black 3-brane is inappropriate when $r_H\sim l_s$, as 
discussed in the Sec.2.

When the horizon size is much larger than the string scale, {\it i.e.,} 
$r_H\gg l_s$, the supergravity description of the interior region is 
valid except very near of the singularity. In this regime, at least in 
gravity side, the phenomenology seems to be very similar to that of the 
conventional large extra dimension scenarios with size $r_H$. 
Since the effective Planck scale is determined via the relation 
$M_{pl}^2=r_H^6/(24\pi^3g_s^2l_s^8)$ and the size of the horizon is
given by $r_H^4=4\pi l_s^4g_sN$, the hierarchy between the string scale 
$l_s^{-1}$ and the four-dimensional Planck scale $M_{pl}$ can be provided 
with the R-R charge of the amount of 
$N=\left(2\pi^{3/2}l_s^2g_s^{1/2}M_{pl}^2\right)^{2/3}$, that is, 
\begin{equation}
\frac{M_{pl}}{l_s^{-1}}\sim N^{3/4},
\end{equation} 
where we have taken the string coupling, $g_s$, to be of the order of one.
If we naively assume that the string scale is of the order of the weak scale, 
{\it i.e.,} $l_s^{-1}\sim m_{EW}$, then the magnitude of the R-R charge needed
to generate the hierarchy is $N\sim 10^{22}$ and then the size of the horizon
is $\sim 10^{-12}$cm. Hence, a D3-brane with large R-R charge, $N\sim 10^{22}$,
provides a setup for a dynamical determination of the large hierarchy between 
the Planck scale and the string scale of the order of the weak scale.  
Moreover, the compactification scale is now determined by the R-R charge of 
the D3-brane and its stabilization is guaranteed by the conserved nature of 
the R-R charge. 

\section{Hawking radiation and small cosmological constant}

Hitherto, we have assumed that our world brane is residing in the interior 
region of an extremal D3-brane. However, it seems that we are living in a 
non-extremal D3-brane rather than the extremal one because the present entropy 
density of our universe is not zero even though it is small, while the extremal 
D3-brane has zero entropy. In this section, we will consider the non-extremal 
version of the D3-brane and briefly discuss a few issues associated to the 
brane world scenarios.

In string frame and with a Schwarzschild-type radial coordinate $r$, 
the metric of the non-extremal D3-brane can be written as \cite{HS}
\begin{equation}\label{nebb}
ds^2=\frac{\Delta_+(r)}{\sqrt{\Delta_-(r)}}\left( -dt^2 +\frac{\Delta_-(r)}
{\Delta_+(r)}d\vec{x}^2\right)
+\frac{dr^2}{\Delta_+(r)\Delta_-(r)}+r^2d\Omega_5^2,
\end{equation}
where $\Delta_{\pm}(r)\equiv 1-r_{\pm}^4/r^4$ and $r_+>r_-$. This metric has a 
nondegenerate, nonsingular outer horizon at $r=r_+$. At $r=r_-$, one encounters
an inner horizon, which, however, coincides with a curvature singularity. 
The singularity is covered with the outer horizon and so it does not matter to 
one who lives outside the outer horizon $r>r_+$. Then it can be regarded as 
a black hole. 

Clearly the metric (\ref{nebb}), however, is appropriate only for $r>r_-$.
A solution which is appropriate for $r<r_-$ can simply be obtained by replacing 
$\Delta_{\pm}(r)$ with $\tilde{\Delta}_{\pm}(r)\equiv r_{\pm}^4/r^4-1$ for 
$r<r_-$ from the metric (\ref{nebb}).
This interior solution interpolates between two singularities at $r=0$ and 
$r=r_-$. The singularity at $r=0$ is due to the $\delta$-function like source 
of the R-R self-dual 5-form field and it could be regarded as a stack of 
D3-branes. While, the singularity at $r=r_-$ is actually deriven by the 
non-linearity of the Einstein equations and disappears in the extremal limit, 
$r_-=r_+=r_H$. It is clear that the interior (the exterior) solution reduce to 
the interior (the exterior) solution of Eq.(\ref{ebs}) in the extremal limit.
The singularity at $r=r_-$ completely disappears as $r_-=r_+$ and the 
Poincar\'{e} invariance in the $x^\mu$-direction is recovered.  

In the non-extremal case, the interior solution is bounded by the naked 
singularity. That is, the curvature singularity sits at some finite proper 
distance from the brane on which visible sector matter exists.
Such naked spacetime singularity may be interpreted as the boundary of the 
extra dimension.\footnote{ 
Similar constructions have appeared in variants of the Randall-Sundrum 
construction \cite{ADKS,KSS,CEGH,FLLN,Luty,CC}. 
And it was proposed that the singularities could help 
solve the cosmological constant problems \cite{ADKS,KSS,CEGH,FLLN,Luty}. 
However, in those constructions a crucial difference from the present one is 
that the geometry with the four-dimensional Poincar\'{e} invariance in the 
$x^\mu$-direction has the naked singularity. While, in the spacetime described
by the metric (\ref{nebb}), the singularity disappears in the limit that the 
Poincar\'{e} invariance in $x^\mu$-direction is recovered.}
The proper interpretation of this singularity will likely be crucial to 
understanding the physics on the world brane. For example, the singularity may 
force the short-distance properties of quantum gravity to become relevant to 
the physics at long distances. 
However, since we do not have complete understandings about the singularity
and the quantum gravity (or stringy effects on such large curvature 
background), we will assume that the singularity is smoothed out by 
some stringy effects. The spacetime then could be extended across the 
singularity into a region where a weakely coupled Einstein decription is valid 
again. There would be various possibilities of the extension across the 
horizon. Here, we will simply assume that the interior region is glued by the 
exterior region of the metric (\ref{nebb}) at the resolved singularity. 
The spacetime then has a causal structure similar to that of the 
non-extremal Reissner-Nordstr\"{o}m black hole. In this case there would be
no 4-dimensional gravity at long distances because the interior region is not
closed and the extra dimensions have an inifinte volume.\footnote{However, we 
need a careful investigation about the role of the spacelike region between
the inner horizon and the outer horizon for the effective 4-dimensional 
gravity.}
However, in the extremal limit the interior region is closed with a finite
volume at the singularity, which is now regular and pushed to infinite proper 
distance. In this case, the extra dimensions are effectively compactified 
as shown in the previous sections, and we should get 4-dimensional gravity
at long distances. 

As mentioned at the beginning of this section, we may live in the interior 
region of a near-extremal D3-brane. Then, how much extremal can the D3-brane be 
it to be our world brane?        
Defining the parameters $r_+$ and $r_-$ as $\delta^4\equiv r_+^4-r_-^4$,
the Hawking temperature and the Bekenstein-Hawking entropy (per unit volume in
the $\vec{x}$-direction) are, respectively, given by
\begin{eqnarray}
T_{\rm H}=\frac{\delta}{\pi r_+^2}~~~{\rm and}~~~ 
s_{BH}\sim\frac{r_+^2\delta^3}{g_s^2l_s^8},
\end{eqnarray} 
up to the numerical factor of the order of one.
The extremal solution has a degenerate horizon $r_+=r_-$, and zero 
Bekenstein-Hawking entropy. The Hawking temperature of the extremal brane 
is also zero. In the near extremal limit, the entropy is  
\begin{eqnarray}
s_{BH}\sim\frac{r_H^{17/4}\epsilon^{3/4}}{g_s^2l_s^8},
\end{eqnarray} 
where $\epsilon\equiv r_+-r_-$. Notice that the black brane entropy is an 
extremely enormous number even for the near extremal black brane with 
$\epsilon\sim l_s$, {\it e.g.,} $s_{BH}\sim 10^{73}/{\rm cm}^3$ as 
$l_s\sim m_{EW}^{-1}$ and $r_H\sim 10^{-12}$cm. On the other hand, the 
present entropy density of our universe is at most $\sim 10^3/{\rm cm}^3$.
This seems to imply that our world brane can easily be embedded in the interior
region of a extremely near-extremal D3-brane. 

Since a non-extremal black brane has nonzero Hawking temperature and Hawking 
radiates, it evolves into an extremal black brane. Notice that the curvature 
singularity at $r=r_-$ is resolved through this process. As discussed in 
Ref.\cite{KMR}, we may take this as a possible mechanism of resolution of 
various problems associated with the brane world scenario. For example, 
for the metric (\ref{nebb}) the bulk curvature leads to violation of the 
$SO(3,1)$ isometry, in the D3-brane worldvolume direction, observed in our 
brane, so there must be some mechanism to flatten the bulk in the absence 
of some symmetry which protect the $SO(3,1)$ isometry \cite{CKR}. 
Clearly, the bulk curvature can be diluted via the Hawking radiation because 
in the extremal limit the bulk curvature vanishes and the $SO(3,1)$ isometry 
of the D3-brane worldvolume is recovered. 

In general, excitations of the extremal black $p$-brane would correspond to
either non-Ricci flat $\hat{g}_{\mu \nu} = \hat{g}_{\mu \nu}(x)$ such that 
$\hat{R}(\hat{g})\neq0$, or even to depend on the extra dimension coordinates 
$\hat{g}_{\mu \nu} = \hat{g}_{\mu \nu}(x, y)$. Then the corresponding 
non-extremal metric would have different forms from the metric Eq.(\ref{nebb}).
Even though we don't have such metric yet, we expect that the spacetime 
structure will be close to that of metric (\ref{nebb}) and so the 
excitations will be diluted via Hawking radiation. Hence, with the same 
argument as in Ref.\cite{KMR} the observed extremely small cosmological 
constant could be explained, provided that our world brane is embedded in 
the interior region of a very near-extremal black brane. The change of the 
vacuum energy density of the brane which may break the Poincar\'{e} invariance
of the D3-brane worldvolume will be diluted through the Hawking radiation 
process. In the same way, the observed flatness and approximate Lorentz 
invariance of our world brane could be also explained.

\section{Conclusions}

We have shown that the interior region inside the horizon of the D3-brane
possesses all of features needed for a realistic brane world scenario which 
interpolates between the Randall-Sundrum scenario and the large extra dimension 
scenario, provided that the singularity at the origin is smoothed out. In this 
picture, the horizon size can be interpreted as the compactification size in 
that the four-dimensional Planck scale $M_{pl}$ is determined by the 
10-dimensional Planck scale $M_*\sim l_P^{-1}=g_s^{-1/4}l_s^{-1}$ and the size
of the horizon $r_H$ via the familiar relation $M_{pl}^2\sim M_*^8 r_H^6$ and 
the effective gravity on a 3-brane residing in the interior region behaves 
as expected in a world with six extra dimensions compactified with size $r_H$.

The most important consequence of the above picture seems that the 
macroscopically large compactification scale can be derived from the underlying
string theory that has only one physical scale, {\it i.e.,} the string scale 
$l_s$. That is, the size of the large extra dimensions is provided by the large 
R-R charge of the D3-brane and its stabilization is strictly guaranteed from 
the charge conservation law. Then, the hierarchy between the Planck scale and 
the string scale is dynamically determined by the magnitude of the R-R charge 
$N$ carried by the D3-brane without introducing any additional scale. 
 
In this picture, the finetuning required in the original RS setup in order to
ensure the vanishing of the four-dimensional cosmological constant could be
eliminated. The vacuum energy density of the brane can be diluted through
the Hawking radiation and the observed extremely small cosmological constant
can be explained provided that our world brane is a very near-extremal 
D3-brane. The observed flatness and approximate Lorentz invariance of our 
world brane could be explained with the same token.


\bigskip\bigskip
\centerline{ {\bf Acknowledgements} }
We would like to thank Sangheon Yi, Jong Dae Park, Hyun Min Lee, O-Kab Kwon
and Soo-Jong Rey for useful discussions. The work is supported in 
part by Korea Research Foundation (KRF-2000-015-BP0072), and by the BK21 
project of Ministry of Education.




\end{document}